\documentclass{elsart4}

\usepackage{graphicx}
\usepackage[T1]{fontenc}
\usepackage{ae}
\usepackage{amssymb}
\usepackage{amsmath}
\usepackage{amsfonts}
\graphicspath{{/Dades/Ciencia/Fe2O3-2002/Js_new(Mn)/EQ_large/},{/Dades/Ciencia/Fe2O3-2002/Js_new(Mn)/HIST/},{/Dades/Ciencia/Fe2O3-2002/Js_new(Mn)/HIST/HMM2003/}}

\begin{document}

\begin{frontmatter}

\title{Influence of surface anisotropy on the hysteresis of magnetic nanoparticles}
%\thanks[tit1]{Title footnote}

\author{\`Oscar Iglesias\corauthref{cor1}}
\ead{oscar@ffn.ub.es}
\ead[url]{http://www.ffn.ub.es/oscar}
%\thanks[label1]{author footnote}
\corauth[cor1]{Tel: 934021155; fax: 934021149}
\author{Am\'{\i}lcar Labarta}
\address{Departament de F\'{\i}sica Fonamental, Universitat de Barcelona, Av. Diagonal 647, 08028 Barcelona, Spain}
%\thanks[label2]{aff footnote}


%\author{}
%\address{}

\begin{abstract}
We present the results of Monte Carlo simulations of the magnetic properties of individual spherical nanoparticles with the aim to explain the role played by surface anisotropy on their low temperature magnetization processes.  
Phase diagrams for the equilibrium configurations have been obtained, showing a change from quasi-uniform magnetization state to a state with hedgehog-like structures at the surface as $k_S$ increases.  
Through the simulated hysteresis loops and the analysis of spin configurations along them, we have identified a change in the magnetization reversal mechanism from quasi-uniform rotation at low $k_S$ values, to a non-uniform switching process at high $k_S$. 
Results for the dependence of the coercive field and remanence on $k_S$ and particle size are also reported.
\end{abstract}

%%%%%%%%%use  the \KEY command at the begin of keyword text%%%%%%%%%
\end{frontmatter}

%%%%%%%%%%%%%%%%%%%%%%%%%%%%%%%%%%%%%%%%%%%%%%%%%%%%%%%%%%%%%%%%%%%%%%%%%%%%%%%%%%%%%%%%%%%%
\section{Introduction}
\label{Intro}
When reducing the size of magnetic particles to the nanoscale, surface and finite-size effects play a dominant role on establishing their peculiar low temperature magnetic properties \cite{Batllejpd02}. Reduction of the magnetic ordering temperature and spontaneous magnetization as particle size is reduced \cite{Mamiyaprl98,Morupprb95}, increased coercivities and high closure fields in the hysteresis loops \cite{Kodamaprb99} are often attributed to the frustration associated to broken bonds and disorder present at the particle surface. 
However, in real samples, these effects are usually masked by the presence of particle size distributions and by the magnetic interactions between the particles (often mentioned as a possible source of their spin-glass-like behavior), making it very difficult to discern what is the real contribution of surface effects alone on the magnetic properties. Moreover, current experimental techniques do not allow to probe surface microscopic structure in geometries such as those of a nanoparticle, a fact that forces to assume phenomenological models for surface anisotropy. Monte Carlo (MC) simulations constitute a benchmark to probe different models since, by modeling magnetic ions as classical \cite{Trohidoujap98,Kachkachiprb02,Labayjap02,Restrepojm04} or Ising spins \cite{Trohidouprb90,Dimitrovprb95,Iglesiasprb01}, the interatomic interactions, the local magnetic anisotropy directions and the values of anisotropy constants can be easily varied while taking into account the exact geometry of the underlying lattice of magnetic ions.  
Here, we will present results of MC simulations aimed to understand the influence of surface anisotropy on the magnetic properties and hysteresis of a single nanoparticle, neglecting the effects of interactions with other particles.
%%%%%%%%%%%%%%%%%%%%%%%%%%%%%%%%%%%%%%%%%%%%%%%%%%%%%%%%%%%%%%%%%%%%%%%%%%%%%%%%%%%%%%%%%%%%
\section{Model and Results}
Since most of the peculiar low temperature magnetic properties of nanoparticles are more clearly observed in ferromagnetic oxides, we will consider here a particle with lattice geometry of maghemite ($\gamma$-Fe$_2$O$_3$), a ferrimagnetic compound in which the magnetic ions sit on two sublattices ($O, T$) with different coordination. We have considered particles with spherical shapes of diameters $D= 3-6 a$ ($a$ is the lattice constant).The $N$ magnetic ions will be modeled by Heisenberg classical spins ${\vec S}_i$ of unit magnitude with the following interaction Hamiltonian in temperature units
\begin{eqnarray}
\label{Hamiltonian}
{ H}/k_{B}= 
-\sum_{\langle  i,j\rangle}J_{ij} {\vec S}_i \cdot {\vec S}_j   
-\sum_{i= 1}^{N} \vec h\cdot{\vec S_i}\nonumber\\
-\sum_{i= 1}^{N_{C}}\left[k_C(S_i^z)^2\right]+\sum_{i= 1}^{N_{S}}\left[k_S(\vec{S}_i \cdot \hat n_i)^2 \right]\ . 	
\end{eqnarray}   
This is an extension to Heisenberg spins of our previous model for Ising spins in which the first term accounts for the n.n. exchange interaction, with $J_{ij}$ the (intra)inter-sublattice (negative) exchange constants (see \cite{Iglesiasprb01} for numerical details) and the second represents the interaction with the external field $\vec h$ (this can be translated to physical units through the relation $\vec H= k_B \vec h /\mu$, with $\mu$ the magnetic moment of Fe ions). The last two terms have been introduced to differentiate the magnetic anisotropy at the $N_C$ core spins, $k_C=1$ (uniaxial along the z-axis), from that at the $N_S$ surface spins (those in the outer unit cell), $k_S$, which will be considered to point along the local radial direction $\hat n_i$ due to breaking of crystalline symmetry at particle boundaries. Our goal is to understand the change in the magnetic properties as $k_S$ is increased from $k_C$ for different particle diameters $D$. Some of the details of the MC method as implemented in our simulations can be found in previous works \cite{IglesiasphaB04,Iglesiasjm04}.
%%%%%%%%%%%%%%%%%%%%%%%%%%  FIG 1 %%%%%%%%%%%%%%%%%%%%%%%%%%%%%%%%%%%%%%%%%%%%%%
\begin{figure}[thbp] 
\centering 
\includegraphics[width=\columnwidth]{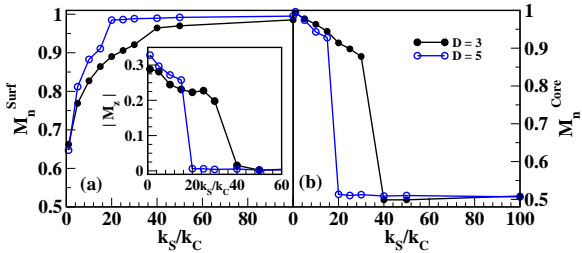}
\caption{(Color online) Phase diagrams showing the contributions of (a) the surface and (b) the core of particles with diameters $D= 3a, 5a$ to the projection along the local easy-axes $M_n$ as a function of the surface anisotropy constant $k_S/k_C$. The Inset in (a) shows the total magnetization along the z-axis $M_z$. 
Below: snapshots of the corresponding $T= 0$ configurations of a spherical particle with $D= 5a$ obtained after cooling from a disordered state at high $T$ are displayed for representative values of $k_S$ in the two regimes. 
}
\label{EQ_Maghemite_Phase_fig}
\end{figure}
%%%%%%%%%%%%%%%%%%%%%%%%%%  FIG 1 %%%%%%%%%%%%%%%%%%%%%%%%%%%%%%%%%%%%%%%%%%%%%%

%\section{Results}

In order to characterize the influence of the surface anisotropy on the magnetic order at low temperatures, we have first performed simulations to find the equilibrium configurations at low $T$. For this purpose, we start from a disordered configuration of spins at a high temperature ($T= 50$K) and follow a progressive cooling down to $T=0$ in constant temperature steps $\delta= -1$K in absence of magnetic field. 
%Thermodynamic averages of the quantities of interest 
%(total magnetization along the z-axis $M_z$, and the corresponding surface and core contributions $M_z^{S}$, $M_z^{C}$) 
%are computed during at least $50000$ MC steps after equilibration during $10000$ MC steps at each $T$. 
In order to better characterize the magnetic ordering of the states achieved at $T=0$, we have also computed the quantity $M_n^{S,C}= \sum_{i=1}^{N_{S,C}} |\vec S_i\cdot \hat n_i|$, which indicates the departure of spins from alignment along the local easy-axis ($M_n=1$ in this case). In Fig. \ref{EQ_Maghemite_Phase_fig}, we present the values of $M_z^{S,C}$ and $M_n^{S,C}$ of the configurations corresponding to $k_S/k_C$ ranging from $1$ to $100$ for particles with diameters $D= 3a, 5a$. The curves in this figure, indicate ferrimagnetic order along the z-axis when $M_n^C\sim 1$ and $M_n^S\sim 0.5$ and radial alignment of spins when $M_n^C\sim 0.5$ and $M_n^S\sim 1$. As can be clearly seen, at a value $k_S = k_S^{\star}$, there is a change in the magnetic order of the configurations as marked by the sharp decrease of $M_n^C$ and $|M_z|$ to zero and the increase of $M_n^S$ to 1. For $k_S< k_S^{\star}$ the particle has a ferrimagnetic ordered core along the z-axis (see left upper snapshot in Fig. \ref{EQ_Maghemite_Phase_fig}) with surface spins that progressively deviate towards the radial direction as $k_S$ increases. This  gives rise to the formation of vortex-like structures that extend to the particle core, as can be seen in the right upper snapshot in Fig. \ref{EQ_Maghemite_Phase_fig}. For $k_S> k_S^{\star}$, the net magnetization of the particle tends towards zero. This is due to the fact that the department of surface spins towards the radial direction is transmitted to the core via the antiferromagnetic exchange interactions, giving rise to hedgehog-like structures (see ower snapshots in Fig. \ref{EQ_Maghemite_Phase_fig}).
%%%%%%%%%%%%%%%%%%%%%%%%%%  FIG 2 %%%%%%%%%%%%%%%%%%%%%%%%%%%%%%%%%%%%%%%%%%%%%%
\begin{figure}[thbp] 
\centering 
\includegraphics[width=\columnwidth]{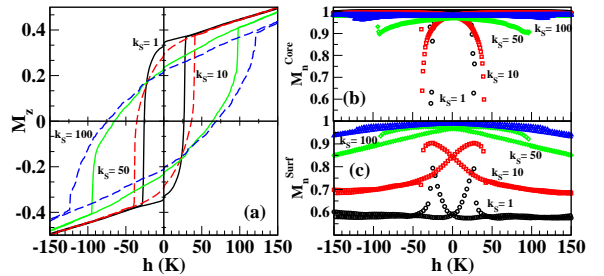}
\caption{(Color online) Hysteresis loops for a particle with $D= 5a$ for different values of $k_S$ as indicated. The different panels stand for (a) the total magnetization, (b) the contribution of the core to $M_n$, and (c) the contribution of the surface to $M_n$.
}
\label{HIST_D5_fig}
\end{figure}
%%%%%%%%%%%%%%%%%%%%%%%%%%  FIG 2 %%%%%%%%%%%%%%%%%%%%%%%%%%%%%%%%%%%%%%%%%%%%%%

\begin{figure}[thbp] 
\centering 
\includegraphics[width=\columnwidth]{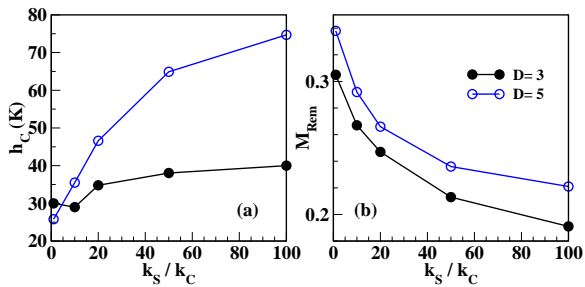}
\caption{(Color online) Dependence of (a) the coercive field $h_C$ and (b) remanent magnetization $M_{Rem}$ on $k_S$ as obtained from low $T$ hysteresis loops for spherical particles with diameters $D= 3a, 5a$.
Lower panels: remanent states along the hysteresis loop of the $D= 5a$ particle at representative values of $k_S$ in the two reversal regimes, snapshots display spin configurations at the equatorial plane perpendicular to the z-axis.
}
\label{Hc(Ks)_Maghemite_fig}
\end{figure}

We have also simulated the hysteresis loops of particles with diameters $D= 3a, 5a$ for different values of $k_S$. We initially apply a strong enough magnetic field along the z-axis strong that is subsequently varied in steps $\delta h= 1$ K 
during which the magnetization is averaged over 1000 MC steps at each field value. 
In Fig. \ref{HIST_D5_fig}, we present hysteresis loops for a particle with $D= 5a$ for representative values of $k_S$. 
Let us first notice that, as surface anisotropy increases, the hysteresis loops change from quasi-squared shape, indicating that the dynamics of the particle may be described as the quasi-uniform rotation of the magnetization due only to the non-compensated spins, to more elongated shape with higher closure fields as observed in oxide particle systems \cite{Martinezprl98}. 
This is in agreement with the formation of disorder states at the particle surface due to increased anisotropy and frustration due to finite-size effects. This change in the reversal mechanism can be more clearly analyzed by comparing panels (b) and (c) in Fig. \ref{HIST_D5_fig}, where, for values of $k_S> k_S^{\star}$, we see that surface spins reverse by forming radial structures ($M_n^S \simeq 1$ along the loop) that induce the subsequent reversal of the core \cite{WWW_JEMS2004}. 
Further insight into the reversal mechanism can be gained from the dependence of the coercive field $h_C$ and remanent magnetization $M_{Rem}$ on $k_S$ presented in Fig. \ref{Hc(Ks)_Maghemite_fig}.
The increase of $h_C$ and the decrease in $M_{Rem}$ observed in Fig.  \ref{Hc(Ks)_Maghemite_fig}, are clear indications of a change in the magnetization reversal mechanism from  quasi-coherent to a surface dominated process \cite{IglesiasphaB04}.
The snapshots of the remanent configurations shown in Fig. \ref{Hc(Ks)_Maghemite_fig} demonstrate the formation of a shell with disordered spins pointing towards the radial direction which increases in thickness as $k_S$ is increased. This last feature also explains the obtained dependence of $h_C$ and $M_{Rem}$ on $k_S$.    
Moreover, $h_C$ increases with decreasing size as observed by Tronc and coworkers in experiments with maghemite particle dispersions \cite{Troncjm03,Troncjm04}.

%%%%%%%%%%%%%%%%%%%%%%%%%%  FIG 3 %%%%%%%%%%%%%%%%%%%%%%%%%%%%%%%%%%%%%%%%%%%%%%
%%%%%%%%%%%%%%%%%%%%%%%%%%  FIG 3 %%%%%%%%%%%%%%%%%%%%%%%%%%%%%%%%%%%%%%%%%%%%%%

%\section{Acknowledgments}
  We acknowledge CESCA and CEPBA under coordination of C$^4$ for computer facilities.
  This work has been supported by the spanish SEEUID through the MAT2003-01124 project.
%and the Generalitat de Catalunya through the 2001SGR00066 CIRIT project.

%\bibliographystyle{prsty}
%{\bibliography{/Dades/Tesi/Refgen,/Dades/Tesi/Refmqt,/Dades/Tesi/Refmqtexp,/Dades/Tesi/Refpart,/Dades/Tesi/Refdip,/Dades/Tesi/Refmc,/Dades/Tesi/Refcamp,/Dades/Tesi/Refdipexp,/Dades/Tesi/Myreferences,/Dades/Tesi/Strings,/Dades/Tesi/Selfassemblies}}

\begin{thebibliography}{99}
\bibitem{Batllejpd02}
X. Batlle and A. Labarta, J. Phys. D: Appl. Phys. {\bf {\bf 35}},  R15  (2002).

\bibitem{Mamiyaprl98}
H. Mamiya, I. Nakatani, and T. Furubayashi, Phys. Rev. Lett. {\bf {\bf 80}},
  177  (1998).

\bibitem{Morupprb95}
S. M{\o}rup, F. B{\o}dker, P.~V. Hendriksen, and S. Linderoth, Phys. Rev. B
  {\bf {\bf 52}},  287  (1995).

\bibitem{Kodamaprb99}
R.~H. Kodama and A.~E. Berkowitz, Phys. Rev. B {\bf {\bf 59}},  6321  (1999).

\bibitem{Trohidoujap98}
K.~N. Trohidou, X. Zianni, and J.~A. Blackman, J. Appl. Phys. {\bf {\bf 84}},
  2795  (1998).

\bibitem{Kachkachiprb02}
H. Kachkachi and M. Dimian, Phys. Rev. B {\bf {\bf 66}},  174419  (2002).

\bibitem{Labayjap02}
Y. Labay {\it et~al.}, J. Appl. Phys. {\bf {\bf 91}},  8715  (2002).

\bibitem{Restrepojm04}
L.~B. J.~Restrepo, Y.~Labaye and J.~M. Greneche, J. Magn. Magn. Mat. {\bf {\bf
  272-276}},  681  (2004).

\bibitem{Trohidouprb90}
K.~N. Trohidou and J.~A. Blackman, Phys. Rev. B {\bf {\bf 41}},  9345  (1990).

\bibitem{Dimitrovprb95}
D.~A. Dimitrov and G.~M. Wysin, Phys. Rev. B {\bf {\bf 51}},  11947  (1995).

\bibitem{Iglesiasprb01}
O. Iglesias and A. Labarta, Phys. Rev. B {\bf {\bf 63}},  184416  (2001).

\bibitem{IglesiasphaB04}
O. Iglesias and A. Labarta, Physica B {\bf {\bf 343}},  286  (2004).

\bibitem{Iglesiasjm04}
O. Iglesias and A. Labarta, J. Magn. Magn. Mat. {\bf {\bf 272-276}},  685
  (2004).

\bibitem{Martinezprl98}
B. Mart\'{\i}nez {\it et~al.}, Phys. Rev. Lett. {\bf {\bf 80}},  181  (1998).

\bibitem{WWW_JEMS2004}
Animated snapshots of the configurations along the hysteresis loops can be
  found at the web site: http://www.ffn.ub.es/oscar/JEMS2004/JEMS2004.html.

\bibitem{Troncjm03}
E. Tronc {\it et~al.}, J. Magn. Magn. Mat. {\bf {\bf 262}},  6  (2003).

\bibitem{Troncjm04}
E. Tronc {\it et~al.}, J. Magn. Magn. Mat. {\bf {\bf 272-276}},  1474  (2004).

\end{thebibliography}

\end{document}